\documentclass[conference]{IEEEtran}
\IEEEoverridecommandlockouts
\usepackage{cite}
\usepackage{amsmath,amssymb,amsfonts}
\DeclareUnicodeCharacter{2009}{\,} 
\usepackage{algorithmic}
\usepackage{graphicx}
\usepackage{hyperref}
\usepackage{textcomp}
\usepackage{xcolor}
\usepackage{mathtools, cuted}
\def\BibTeX{{\rm B\kern-.05em{\sc i\kern-.025em b}\kern-.08em
    T\kern-.1667em\lower.7ex\hbox{E}\kern-.125emX}}

\makeatletter
\def\ps@IEEEtitlepagestyle{%
  \def\@oddfoot{\mycopyrightnotice}%
  \def\@evenfoot{}%
}
\def\mycopyrightnotice{%
  \begin{minipage}{\linewidth}
  \scriptsize
  XXX-X-XXXX-XXXX-X/XX/\$31.00~\copyright~2024 IEEE
  \end{minipage}
}

\newcommand{\linebreakand}{%
  \end{@IEEEauthorhalign}
  \hfill\mbox{}\par
  \mbox{}\hfill\begin{@IEEEauthorhalign}
}
\let\old@ps@IEEEtitlepagestyle\ps@IEEEtitlepagestyle
\def\confheader#1{%
    \def\ps@IEEEtitlepagestyle{%
        \old@ps@IEEEtitlepagestyle%
        \def\@oddhead{\strut\hfill#1\hfill\strut}%
        \def\@evenhead{\strut\hfill#1\hfill\strut}%
    }%
    \ps@headings%
}
\makeatother

\begin{document}%

\title{\fontsize{24}{8}\selectfont The Zeta Notation ($\zeta$) for Complex Asymptotes
}
%
%
\author{\IEEEauthorblockN{\fontsize{9}{8}\selectfont Anurag Dutta}
\IEEEauthorblockA{\textit{\fontsize{9}{8}\selectfont Department of Computer Science \& Engineering} \\
\textit{\fontsize{9}{8}\selectfont Government College of Engineering \& Textile Technology}\\
\fontsize{9}{8}\selectfont Serampore, India \\
\fontsize{9}{8}\selectfont anuragdutta@ieee.org}
\and
\IEEEauthorblockN{\fontsize{9}{8}\selectfont Lakshmanan Kumarasankaralingam}
\IEEEauthorblockA{\textit{\fontsize{9}{8}\selectfont Department of Mathematics} \\
\textit{\fontsize{9}{8}\selectfont St. Joseph University, Virgin Town}\\
\fontsize{9}{8}\selectfont Nagaland, India\\
\fontsize{9}{8}\selectfont lakshmanlingam@kas.edu.kw}
\linebreakand
\IEEEauthorblockN{\fontsize{9}{8}\selectfont John Harshith}
\IEEEauthorblockA{\textit{\fontsize{9}{8}\selectfont Department of Computer Science \& Engineering} \\
\textit{\fontsize{9}{8}\selectfont Vellore Institute of Technology}\\
\fontsize{9}{8}\selectfont Vellore, India\\
\fontsize{9}{8}\selectfont john.harshith2020@vitstudent.ac.in}
\and
\IEEEauthorblockN{\fontsize{9}{8}\selectfont Athilingam Ramamoorthy}
\IEEEauthorblockA{\textit{\fontsize{9}{8}\selectfont Department of Mathematics} \\
\textit{\fontsize{9}{8}\selectfont Velammal Engineering College}\\
\fontsize{9}{8}\selectfont Chennai, India\\
\fontsize{9}{8}\selectfont ramamoorthy@velammal.edu.in}
\linebreakand
\IEEEauthorblockN{\fontsize{9}{8}\selectfont ~~~~~~~~Chandrasekar Pradeep}
\IEEEauthorblockA{\textit{\fontsize{9}{8}\selectfont ~~~~~~~~Department of Mathematics} \\
\textit{\fontsize{9}{8}\selectfont ~~~~~~~~MVJ College of Engineering, Bangalore}\\
\fontsize{9}{8}\selectfont ~~~~~~~~Karnataka, India\\
\fontsize{9}{8}\selectfont ~~~~~~~~e.chandrasekarpradeep@gmail.com}
\and
\IEEEauthorblockN{\fontsize{9}{8}\selectfont ~~~~~~Pijush Kanti Kumar}
\IEEEauthorblockA{\textit{\fontsize{9}{8}\selectfont ~~~~~~Department of Information Technology} \\
\textit{\fontsize{9}{8}\selectfont ~~~~~~Govt. College of Engg. \& Textile Tech.}\\
\fontsize{9}{8}\selectfont ~~~~~~Serampore, India \\
\fontsize{9}{8}\selectfont ~~~~~~pijush752000@yahoo.com}
}

\maketitle              
\begin{abstract}
Time Complexity is an important metric to compare algorithms based on their cardinality. The commonly used, trivial notations to qualify the same are the Big-Oh, Big-Omega, Big-Theta, Small-Oh, and Small-Omega Notations. All of them, consider time a part of the real entity, i.e., Time coincides with the horizontal axis in the argand plane. But what if the Time rather than completely coinciding with the real axis of the argand plane, makes some angle with it? We are trying to focus on the case when the Time Complexity will have both real and imaginary components. For Instance, if $T\left(n\right)=\ n\log{n}$, the existing asymptomatic notations are capable of handling that in real time But, if we come across a problem where, $T\left(n\right)=\ n\log{n}+i\cdot n^2$, where, $i=\sqrt[2]{-1}$, the existing asymptomatic notations will not be able to catch up. To mitigate the same, in this research, we would consider proposing the Zeta Notation ($\zeta$), which would qualify Time in both the Real and Imaginary Axis, as per the Argand Plane. \\
\end{abstract}


\section{Introduction}\label{sec1}
Computing complexity theory within speculative mathematics and computer science is concerned with categorizing computation issues according to respective resource requirements and connecting different categories. Every computing task is a challenge that a computer can handle. The physical application of arithmetic computations, for instance, an algorithm~\cite{ref_article1}, can solve a computation task. The evaluation of strategies and tractability theories~\cite{ref_article2} are subfields in mathematical computer science that are strongly linked. The aforementioned is completely dedicated to analyzing the number of resources required by a proposed methodology to come up with a solution, whereas the latter starts asking a rather more straightforward question about certain possible methodologies that could be employed to address the same issue. This is a critical distinction between both the assessment of methodologies and the complexity of the algorithm theory. The classification of problems that can or cannot be addressed with adequately constrained resources is the aim of computational complexity research. The distinction between computational effort and computability theory, which examines what types of problems may, in general, be solved programmatically, is the imposition of constraints~\cite{ref_article3} on the accessible resources. The terms "best case complexities," "worst case complexity," and "average case complexity"~\cite{ref_article4} relate to three distinct methods of calculating the computational complexity or any additional type of comportment of several contributions of identical size. We stipulate the following complexity since some inputs of size $n$ may be more straightforward to solve than others:-
\begin{enumerate}
    \item \textit{Best-Case Complexity}: The difficulty of addressing the issue for the ideal intake of size $n$ is represented by this.
    \item \textit{Average-Case Complexity}: On average, this is how difficult the problem is to solve. Only a probability density function across the inputs is taken into consideration while defining this complexity. For instance, the average case intricacy can be calculated according to the uniform distribution across all entries of size $n$ if it is assumed that all entries of the same size are equally likely to appear.
    \item \textit{Amortized Analysis}: In an amortized estimate, all of the computation operations both expensive as well as inexpensive are taken into account.
    \item \textit{Worst-Case Complexity}: The difficulty of addressing the issue for the worst possible intake of size $n$ is represented by this.
\end{enumerate}
It is useful to show upper and lower limits mostly on the greatest amount of time necessitated from the most effective procedure to tackle a particular situation to categorize the computational difficulties. Unless otherwise specified, the cost of an approach is typically assumed to represent its worst-case complication. The discipline of algorithm analysis includes analyzing a specific algorithm. One just needs to demonstrate the existence of a specific algorithm with an execution time of no more than $T(n)$ to demonstrate an optimum value $T(n)$ on the temporal hardness~\cite{ref_article5} of a task. Lower bounds~\cite{ref_article6}, on the other hand, are significantly harder to demonstrate because they generalize all potential solutions to a specific problem. The statement "all potential strategies" refers to not only the techniques that are generally recognized but also any prospective computational methods that may be developed. It is necessary to demonstrate that no methodology can have a time complexity~\cite{ref_article1} fewer than $T(n)$ to provide a lower constraint of $T(n)$ for a subject. 
\begin{figure}[htbp]
\includegraphics[width=\linewidth]{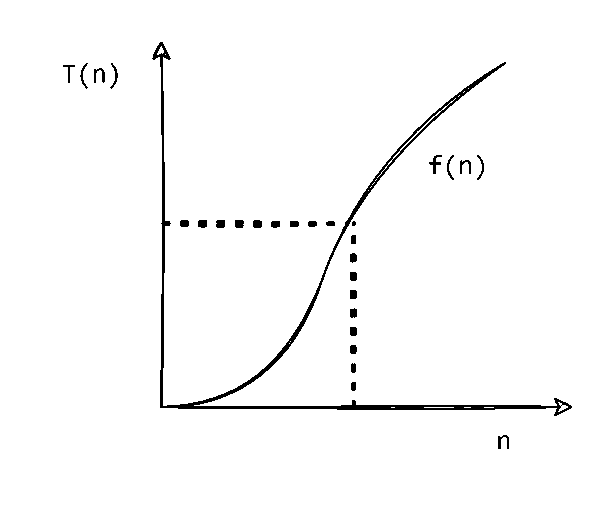}
\caption{Graphical Implication of the Real valued asymptotic notations. The ordinates here stand for the computational time, while the abscissae implicate the cardinally of the data input.} \label{fig1}
\end{figure}
Here, it’s quite observable that the General or prevalent asymptotic notations~\cite{ref_article8} are of Real Order (Figure \ref{fig1}). They don’t take into account the complexities of those in terms of iota ($i$). In this paper, we would propose, $\zeta$ Notation, which is supposed to take care of the Imaginary Computational Complexities. Further, we would propose a transformation that could project Computational Complexity in Generalized Asymptotic Notations to the Notations in Imaginary Domains. Section \ref{sec2} proposes the $\zeta$ Notation. Section \ref{sec3} proposes the transformation and some necessary conversions needed for the smooth transformation of the modern-day Computational Complexities in Real Domains to the Imaginary Domains taking into account the Zeta Notation. Section \ref{sec4} poses a Comparative Analysis with the pre-existing Notations, while Section \ref{sec5} summarizes the research.
\section{The Zeta ($\zeta$) Notation}\label{sec2}
\begin{figure}[htbp]
\includegraphics[width=\linewidth]{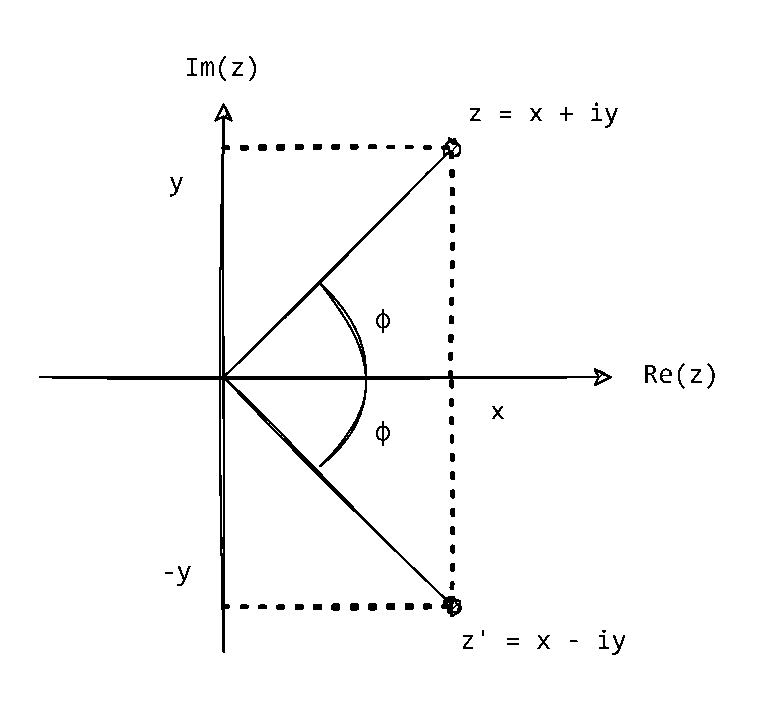}
\caption{An example of a complex plane. Upon this vertical spatial alliance are just the imaginary numbers. On the Horizontal spatial alliance, are the real numbers. The angle $\phi$ is the one that the Complex Number ($z$) subtends on the Real Axes. According to this illustration, multiplying by -1 causes a half circle's origin to rotate $180^o$. A spin of $90^o$ around the center, which corresponds to a quarter of a circle, results from multiplying by $i$.} \label{fig2}
\end{figure}

Any true figure being multiplied by such an imaginary~\cite{ref_article9} component, $i$ whose feature $i^2=-1$ defines it as an imaginary number. Real numbers $a$ and $b$ are referred to, correspondingly, as the real portion and the imaginary portion of the complex number when they are combined with an imaginary number $b\cdot i$ to create a complex number of the form $a+b\cdot i$ (Figure \ref{fig2}). 
Computational Complexities in the current practical world deal with Generalized Asymptotic Notations like Big-Oh Notation~\cite{ref_article10}, Big-Omega Notation~\cite{ref_article11}, Big-Theta Notation~\cite{ref_article12}, none takes into account, the Imaginary Plane, better known as the Argand Plane. The complex plane in mathematics is the plane made up of complex numbers, with real and imaginary numbers constituting the x-axis and real and imaginary, respectively, of the y-axis in a Cartesian coordinate system. Complex numbers can be geometrically interpreted using the complex plane. They add feature vectors when addition is applied. Polar coordinates make it simpler to represent the product of two complex numbers; the magnitude, or modulus, of the products, is the sum of both of the angles, or arguments, and the product's angle, or argument, is the product of its two actual values, or moduli. In specifically, rotation is produced when modulus 1 is multiplied by a complex number. The Argand plane~\cite{ref_article13} as well as the Gauss plane are alternate names for the complicated plane. The $\zeta$ Notation will be based on the Argand Plane with its axes being, the Real Part of the Computational Complexity, the Imaginary Part of the Computational Complexity, and the cardinality~\cite{ref_article14} of the dataset, we want to work upon.
\begin{figure}[htbp]
\includegraphics[width=\linewidth]{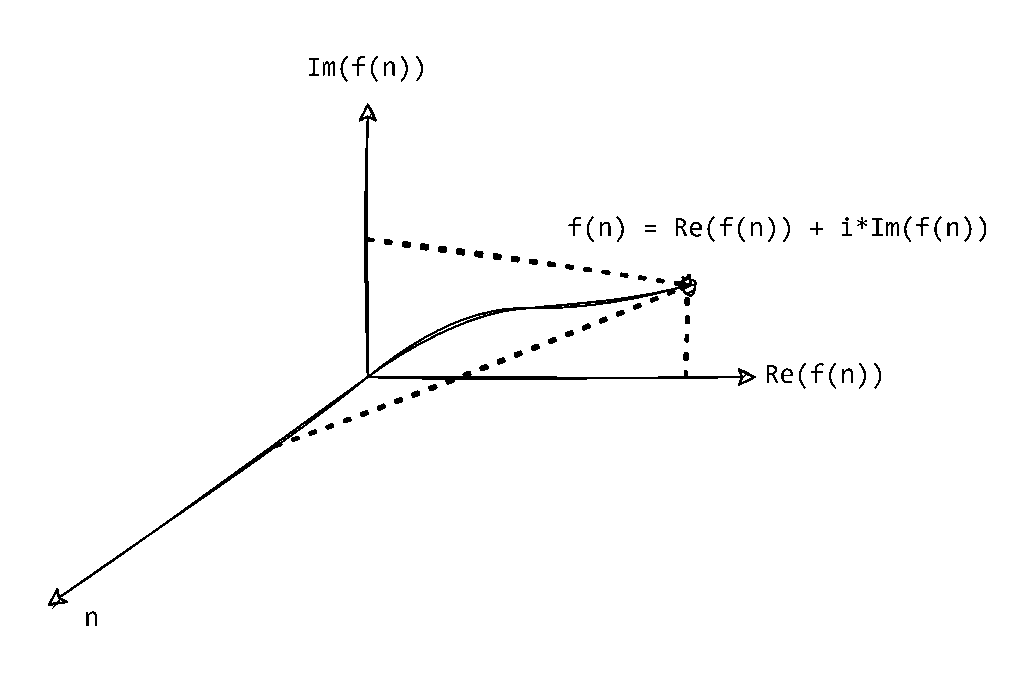}
\caption{Graphical Representation of the Computational Complexity with consideration of the Imaginary Part along with the cardinality of the Dataset.} \label{fig3}
\end{figure}
As stated previously, the Zeta Notation will consider, the imaginary computational entities as well. So, we can represent the Computational Complexity asymptotically, as, 
\begin{equation}
    \label{eq1}
    T(n)=\zeta\left(f(n)\right)=\text{Re}\left(f(n)\right)+i\cdot \text{Im}\left(f(n)\right)
\end{equation}
where, $i=\sqrt[2]{-1}$
Figure \ref{fig4} represents a projection~\cite{ref_article15} of the Complexity of Computational Function, $f(n)$ in 2D Argand (see Eq$^n$ \ref{eq1}). 
\begin{figure}[htbp]
\includegraphics[width=\linewidth]{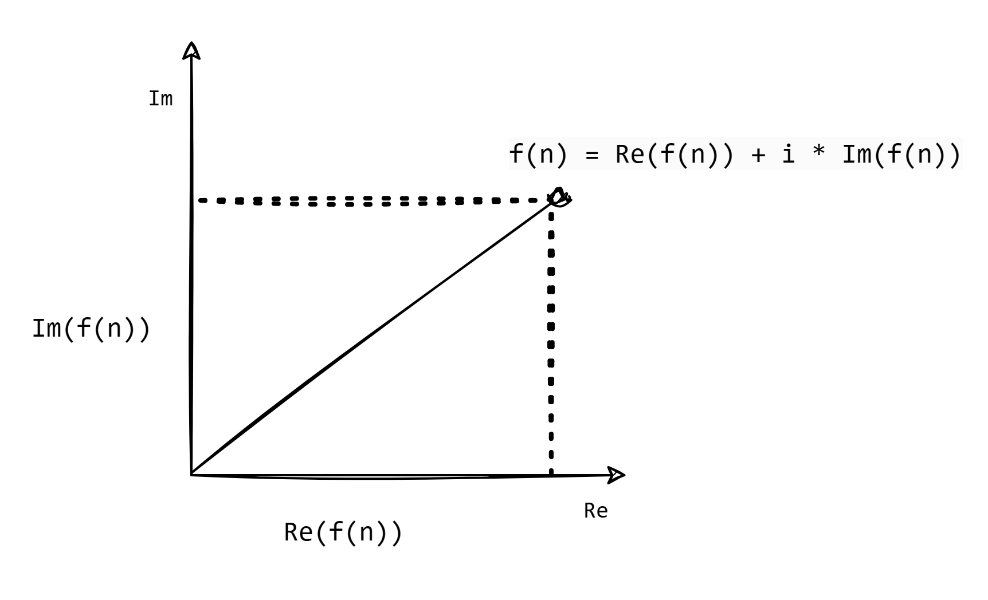}
\caption{Graphical Representation of the Computational Complexity before being spitted in terms of its trigonometric components in a bi-dimensional plane.} \label{fig4}
\end{figure}
Now, if we deep dive into the Asymptotes~\cite{ref_article16}, according to Figure \ref{fig3}, the Computational Complexity Function, $f(n)$ will be evident in the $z-x$ plane. Elsewhere, $f(n)=\lambda=Constant$. 
The same can be represented using the Euler’s Represent~\cite{ref_article17} of the Complex Numbers. 
According to Euler, 
\begin{equation}
    \label{eq2}
    \text{Re}\left(f(n)\right)+i\cdot \text{Im}\left(f(n)\right)=\lvert f(n) \rvert \cdot e^{{i\cdot \text{tan}}^{-1}{\left(\frac{\text{Im}\left(f(n)\right)}{\text{Re}\left(f(n)\right)}\right)}}
\end{equation}
where, $\lvert f(n) \rvert = \left(\sqrt[2]{\left(\text{Re}\left(f(n)\right)\right)^2+\left(\text{Im}\left(f(n)\right)\right)^2}\right)$
Therefore, representation of \ref{eq1} according to Eq$^n$ \ref{eq2}, would be $T(n)=\zeta\left(g(n),\ \phi\right)=e^{{i\cdot \text{tan}}^{-1}{\left(\frac{\text{Im}\left(f(n)\right)}{\text{Re}\left(f(n)\right)}\right)}}\cdot\left(\sqrt[2]{\left(\text{Re}\left(f(n)\right)\right)^2+\left(\text{Im}\left(f(n)\right)\right)^2}\right)$, 
is the main motivation behind the $\zeta$ Notation, where, \\
$g(n)=\sqrt[2]{\left(\text{Re}\left(f(n)\right)\right)^2+\left(\text{Im}\left(f(n)\right)\right)^2}$ and $\phi={\text{tan}}^{-1}{\left(\frac{\text{Im}\left(f(n)\right)}{\text{Re}\left(f(n)\right)}\right)}$.
The Euler implication can be best demonstrated in the Polar Coordinate System~\cite{ref_article18}. 
\begin{figure}[]
\includegraphics[width=\linewidth]{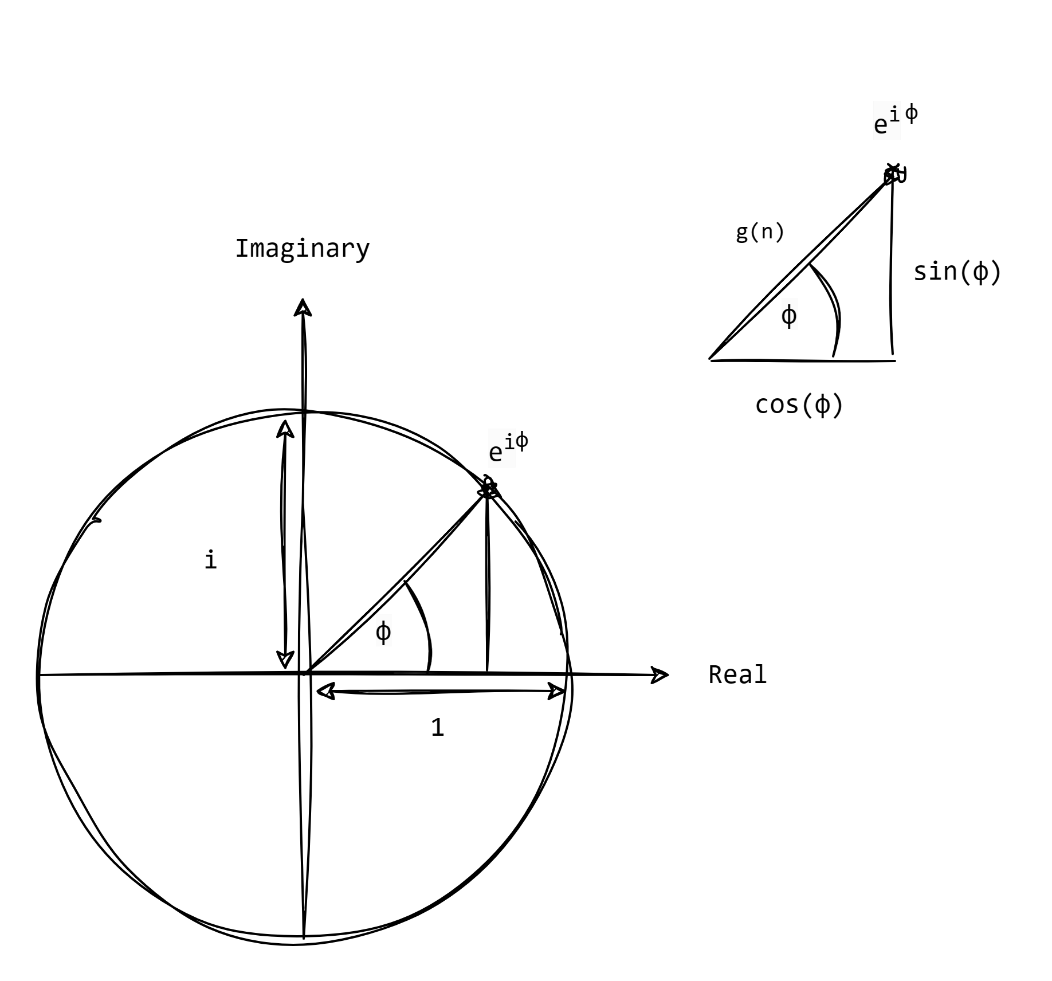}
\caption{Pictorial Representation of the $\zeta$ Notation on the Polar Coordinate System. The circle is having a radial length of $\sqrt[2]{\left(\text{Re}\left(f(n)\right)\right)^2+\left(\text{Im}\left(f(n)\right)\right)^2}$ and will subtend an angle of ${\text{tan}}^{-1}{\left(\frac{\text{Im}\left(f(n)\right)}{\text{Re}\left(f(n)\right)}\right)}$ with the positive $x$ axes. } \label{fig5}
\end{figure}
\section{Transformations to the General Asymptotes}\label{sec3}
\begin{figure}[h]
\includegraphics[width=\linewidth]{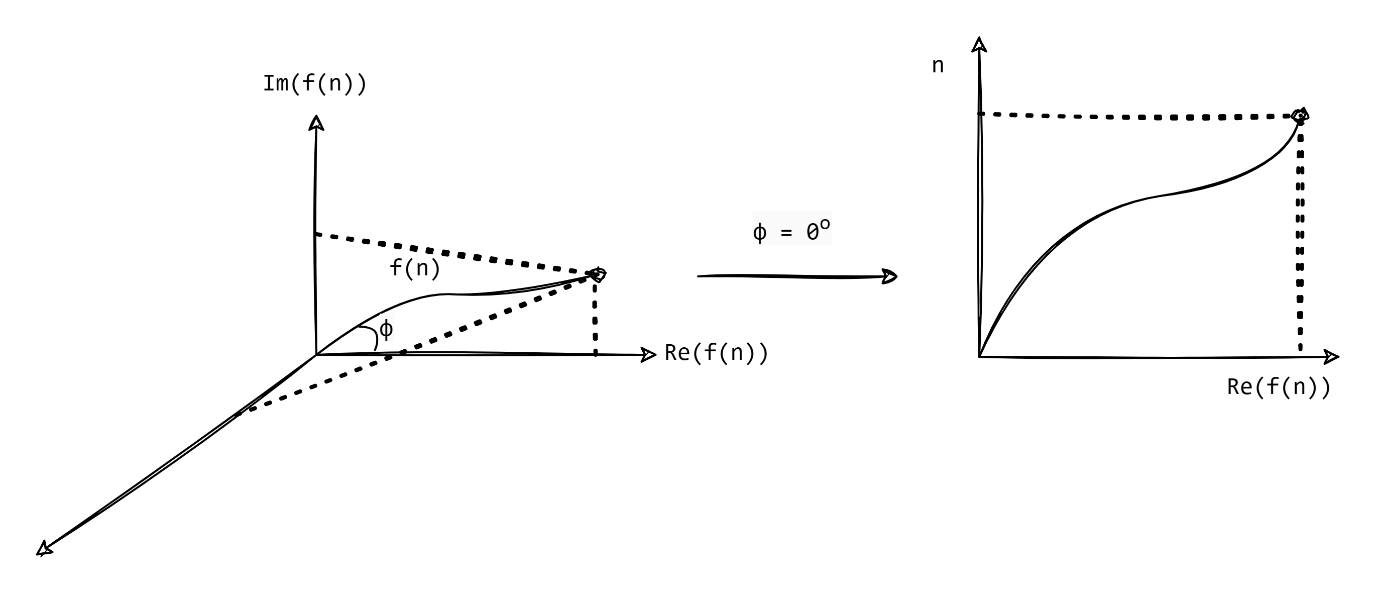}
\caption{Pictorial Implication of the Transformation of Zeta Notation to Generalized Real Domain Computational Complexity. } \label{fig6}
\end{figure}

Now, not always, we need $\zeta$ Notation Asymptotes. Many times, we will find Computational Complexities in Real domains, defying their existence in the Imaginary Domains. The $\zeta$ Notation is capable of smooth transformation to the Generalized Asymptotes, assisted by a shift in origin. In general, Asymptotic Notations, say for instance, a Big Oh, or Omega, or Theta, or a general one, $\varrho(.)$ is related to the computational complexity as, 
\begin{equation}
    \label{eq4}
    T_1(n)=\varrho(f(n))=\alpha \cdot f(n)+\beta
\end{equation}
In contrast, the Zeta Notation relates the Computational Complexity as, 
\begin{equation}
    \label{eq5}
    T_2(n)=\zeta(g(n))=e^{i\phi}\cdot g(n)=(cos\left(\phi\right)+i\cdot sin\left(\phi\right))\cdot g(n)
\end{equation}
The transformation~\cite{ref_article19} would be possible only if the imaginary part of the Computational Complexity is 0. Refraining from that, $sin\left(\phi\right)=0$. Thus, accordingly, $cos\left(\phi\right)=1$.\\
Letting, $T_1(n)=T_2(n)$ (see Eq$^n$ \ref{eq4} and \ref{eq5}) to equate them, $f(n)=\frac{g(n)-\beta}{\alpha}$.\\
Thus, the transformation (Figure \ref{fig6}), would be, 
\begin{equation}
    \varrho\left(\frac{g(n)-\beta}{\alpha}\right)=\zeta\left(g(n)\right)\ \forall g(n):T_1(n)\rightarrow T_2(n)
\end{equation}
where $\varrho$ is the Generalized Asymptotic Notation. 
\section{Comparative Analysis}\label{sec4}
Here, in this section, we will pull a Comparative Analysis between the $\zeta$ Notation and the generalized Asymptotic Notations. 
Big O notation is a type of mathematical notation that expresses how a function limits itself when the parameter tends to zero or infinity. Big O notation in computing science is employed to categorize algorithms based on how their performance or storage needs increase as the number of inputs climbs. Formally, 
\begin{equation}
    \psi(x)=O\left(f\left(x\right)\right)\ni x\rightarrow\infty
\end{equation}
if, for any and all reasonably large values of $x$, the magnitude of $\psi(x)$ is not more than a non-negative multiple of $f\left(x\right)$. Various functions that have identical terminal paces of growth might well be written employing the same O nomenclature. Big O representation describes functions thus according to respective growth rates. The reason the rate of growth of a curve is indeed known as the ordering of the function is why the character O is used. Big O notation often only gives an arbitrary cap on the function's pace of growth when describing a curve. But, Big – Oh notation is unable to handle if $f\left(x\right)$ brings with it, an imaginary component.\\ 
Another mathematical notation is the Big $\Omega$ notation. It expresses how a function limits itself when the parameter tends to some real number. Formally, 
\begin{equation}
    \psi(x)=\Omega\left(f\left(x\right)\right)\ni x\rightarrow a\ \forall\ a\in\mathbb{R}
\end{equation}
$\psi(x)$ and $f\left(x\right)$ are defined in the neighbour of $a$, and $f\left(x\right)$ is non negative in the neighbourhood. But again, this notation is unable to draw about a function having imaginary touch. \\
One more commonly employed mathematical notation is the Big $\Theta$ notation. Formally, 
\begin{equation}
    \psi(x)=\Theta\left(f\left(x\right)\right)
\end{equation}
$\psi(x)$ is sandwiched both above and below by $f\left(x\right)$ asymptotically. But even it is unable to touch the imaginaries. 
\section{Conclusion}\label{sec5}
The Zeta notation is supposed to take the imaginary or theoretically explainable Computational Complexities into account. The $\zeta$ notation itself can be easily transformed to Generalized Asymptotic Notations using a basic One-Degree-Transformation. This notation would help the algorithmists to get a touch on the Imaginary Computational Complexities in vision. That would also help them to visualize the algorithms in a much better way. Many Computational Complexities are such that, they combine themselves up with the Complex Numbers (specifically, $i$). As of now, we don’t have a Computational Asymptotic Notation that could account for those, but the $\zeta$ notation is going to schlep those Computational Complexities to act them up in the Complex Domains. \\

Some, scopes of further research include improvisation of the $\zeta$ notation making use of Complex Analysis. This novel Asymptotic Notation is devoid of any special hardware requirements. The Notation could also be understood well theoretically, without any major technological intervention.

\end{document}